\documentclass{article} 

\usepackage{amssymb,stmaryrd}

\newcommand{\fa}{\forall}
\newcommand{\ex}{\exists}
\newcommand{\topc}{\top^c}
\newcommand{\botc}{\bot^c}
\newcommand{\negc}{\neg^c}
\newcommand{\wedgec}{\wedge^c}
\newcommand{\veec}{\vee^c}
\newcommand{\Rightarrowc}{\Rightarrow^c}
\newcommand{\fac}{\fa^c}
\newcommand{\exc}{\ex^c}

\newtheorem{definition}{Definition}
\newtheorem{proposition}{Proposition}
\newtheorem{theorem}{Theorem}
\newtheorem{corollary}{Corollary}

\newbox\tempa
\newbox\tempb
\newdimen\tempc
\def\mud#1{\hfil $\displaystyle{\mathstrut #1}$\hfil}
\def\rig#1{\hfil $\displaystyle{#1}$}
\def\irulehelp#1#2#3{\setbox\tempa=\hbox{$\displaystyle{\mathstrut #2}$}%
                        \setbox\tempb=\vbox{\halign{##\cr
        \mud{#1}\cr
        \noalign{\vskip\the\lineskip}
        \noalign{\hrule height 0pt}
        \rig{\vbox to 0pt{\vss\hbox to 0pt{${\; #3}$\hss}\vss}}\cr
        \noalign{\hrule}
        \noalign{\vskip\the\lineskip}
        \mud{\copy\tempa}\cr}}
                      \tempc=\wd\tempb
                      \advance\tempc by \wd\tempa
                      \divide\tempc by 2 }
\def\irule#1#2#3{{\irulehelp{#1}{#2}{#3}
                     \hbox to \wd\tempa{\hss \box\tempb \hss}}}

\begin{document}
\title{On the definition of the classical\\ connectives and quantifiers}
\author{Gilles Dowek\thanks{INRIA, 
23 avenue d'Italie, CS 81321, 75214 Paris Cedex 
13, France.
{\tt gilles.dowek@inria.fr}.}}
\date{}

\maketitle
\thispagestyle{empty}

\begin{abstract}
Classical logic is embedded into constructive logic, through a 
definition of the classical connectives and quantifiers in terms 
of the constructive ones.
\end{abstract}

The history of the notion of constructivity started with a dispute on
the deduction rules that one should or should not use to prove a
theorem. Depending on the rules accepted by the ones and the others,
the proposition $P \vee \neg P$, for instance, had a proof or not.

A less controversial situation was reached with a classification of
proofs, and it became possible to agree that this proposition had a
classical proof but no constructive proof.

An alternative is to use the idea of Hilbert and Poincar\'e that
axioms and deduction rules define the meaning of the symbols of the
language and it is then possible to explain that some judge the
proposition $P \vee \neg P$ true and others do not because they do not
assign the same meaning to the symbols $\vee$, $\neg$, etc. The need
to distinguish several meanings of a common word is usual in
mathematics. For instance the proposition ``there exists a number $x$
such that $2 x = 1$'' is true of false depending on whether the word
``number'' means ``natural number'' or ``real number''. Even for
logical connectives, the word ``or'' has to be disambiguated into
inclusive and exclusive.

Taking this idea seriously, we should not say that the proposition $P
\vee \neg P$ has a classical proof but no constructive proof, but we
should say that the proposition $P \veec \negc P$ has a proof and 
the proposition $P \vee \neg P$ does not, that is we should introduce
two symbols for each connective and quantifier, for instance a symbol
$\vee$ for the constructive disjunction and a symbol $\veec$ for the
classical one, instead of introducing two judgments: ``has a
classical proof'' and ``has a constructive proof''.  We should also be
able to address the question of the provability of mixed propositions
and, for instance, express that the proposition $(\neg (P \wedge Q))
\Rightarrow (\neg P \veec \neg Q)$ has a proof.

The idea that the meaning of connectives and quantifiers is expressed
by the deduction rules leads to propose a logic containing all the
constructive and classical connectives and quantifiers and deduction
rules such that a proposition containing only constructive connectives
and quantifiers
has a proof in this logic if and only if it has a proof in
constructive logic and a proposition containing only classical
connectives and quantifiers
has a proof in this logic if and only if it has a proof in
classical logic.  Such a logic containing classical, constructive, and
also linear, connectives and quantifiers has been proposed by J.-Y. Girard
\cite{Girard}. This logic is a sequent calculus with {\em unified
  sequents} that contain a linear zone and a classical zone and rules
treating differently propositions depending on the zone they belong.

Our goal in this paper is slightly different, as we want to define the
meaning of a small set of primitive connectives and quantifiers with
deduction rules and define the others explicitly, in the same way the
exclusive or is explicitly defined in terms of conjunction,
disjunction and negation: $A \oplus B = (A \wedge \neg B) \vee (\neg A
\wedge B)$.  A first step in this direction has been made by G\"odel
\cite{GodelS4} who defined a translation of constructive logic into
classical logic, and Kolmogorov \cite{Kolmogorov}, G\"odel
\cite{Godel}, and Gentzen \cite{Gentzen} who defined a
translation of classical logic into constructive logic.
As the first translation requires a modal
operator, we shall focus on the second. This leads to consider
constructive connectives and quantifiers as primitive and search for
definitions of the classical ones.  Thus, we want to define classical
connectives and quantifiers $\topc$, $\botc$, $\negc$, $\wedgec$,
$\veec$, $\Rightarrowc$, $\fac$, and $\exc$ and embed classical
propositions into constructive logic with a function $\|~\|$ defined
as follows.
\begin{definition}~
\begin{itemize}
\item $\|P\| = P$ if $P$ is an atomic proposition
\item $\|\top\| = \topc$
\item $\|\bot\| = \botc$
\item $\|\neg A\| = \negc \|A\|$
\item $\|A \wedge B\| = \|A\| \wedgec \|B\|$
\item $\|A \vee B\| = \|A\| \veec \|B\|$
\item $\|A \Rightarrow B\| = \|A\| \Rightarrowc \|B\|$
\item $\|\fa x~A\| = \fac x~\|A\|$
\item $\|\ex x~A\| = \exc x~\|A\|$
\end{itemize}
If $\Gamma = A_1, ..., A_n$ is a multiset of propositions, we write 
$\|\Gamma\|$ for the multiset $\|A_1\|, ..., \|A_n\|$.
\end{definition}

Kolmogorov-G\"odel-Gentzen translation can be defined as follows
\begin{itemize}
\item $(P)' = \neg \neg P$, if $P$ is an atomic proposition
\item $(\top)' = \neg \neg \top$ 
\item $(\bot)' = \neg \neg \bot$ 
\item $(\neg A)' = \neg \neg \neg (A)'$
\item $(A \wedge B)' = \neg \neg ((A)' \wedge (B)')$
\item $(A \vee B)' = \neg \neg ((A)' \vee (B)')$
\item $(A \Rightarrow B)' = \neg \neg ((A)' \Rightarrow (B)')$
\item $(\fa x~A)' = \neg \neg (\fa x~(A)')$
\item $(\ex x~A)' = \neg \neg (\ex x~(A)')$
\end{itemize}
or more succinctly as 
\begin{itemize}
\item $(P)' = \neg \neg P$, if $P$ is an atomic proposition
\item $(*)' = \neg \neg *$, if $*$ is a zero-ary connective
\item $(* A)' = \neg \neg (* (A)')$, if $*$ is a unary connective
\item $(A * B)' = \neg \neg ((A)' * (B)')$, if $*$ is a binary connective
\item $(*x~A)' = \neg \neg (*x~(A)')$, if $*$ is a quantifier
\end{itemize}
For instance 
$$(P \vee \neg P)' = 
\neg \neg (\neg \neg P \vee \neg \neg \neg \neg \neg P)$$
And it is routine to prove that 
a proposition $A$ has a classical proof if and only if the
proposition $(A)'$ has a constructive one.  

But, this translation does not exactly provide a definition of the
classical connectives and quantifiers in terms of the constructive
ones, because an atomic proposition is $P$ is translated as $\neg \neg
P$, while in a translation induced by a definition of the classical
connective and quantifiers, an atomic proposition $P$ must be
translated as $P$.

Thus, to view Kolmogorov-G\"odel-Gentzen translation as a definition,
we would need to also introduce a proposition symbol $P^c$ defined by
$P^c = \neg \neg P$. But this would lead us too far: we want to
introduce constructive and classical versions of the logical
symbols---the connectives and the quantifiers---but not of the non
logical ones, such as the predicate symbols.

If we take the definition 
\begin{itemize}
\item $\negc A = \neg \neg \neg A$
\item $A \veec B = \neg \neg (A \vee B)$
\item etc.
\end{itemize}
where a double negation is put before each connective and
quantifier, then the proposition $P \veec \negc P$ is $\neg \neg
(P \vee \neg \neg \neg P)$ where, compared to 
the Kolmogorov-G\"odel-Gentzen translation, 
the double
negations in front of atomic propositions are missing.  
Another translation introduced by L. Allali and O. Hermant \cite{AH}
leads to the definition 
\begin{itemize}
\item $\negc A = \neg \neg \neg A$
\item $A \veec B = (\neg \neg A) \vee (\neg \neg B)$
\item etc.
\end{itemize}
where double negations are put after, and not before, each connective and
quantifier.
The proposition $P \veec \negc P$ is then $\neg \neg P \vee \neg \neg \neg \neg \neg P$,
where the double negation at the top of the proposition is missing.
Using this translation Allali and Hermant prove that the proposition
$A$ has a classical proof if and only if the proposition $\neg \neg
\|A\|$ has a constructive one and they introduce another provability
judgment expressing that the proposition $\neg \neg A$ has a
constructive proof. This also would lead us too far: in our logic, 
we want a single
judgment ``$A$ has a proof'' expressing that $A$ has a constructive
proof, and not to introduce a second judgment, whether it be ``$A$ has
a classical proof'' or ``$\neg \neg A$ has a proof''.

In order to do so, we define the classical connectives and quantifiers by
introducing double negations both before and after each symbol.
\begin{definition}[Classical connectives and quantifiers]~
\begin{itemize}
\item $\topc = \neg \neg \top$
\item $\botc = \neg \neg \bot$
\item $\negc A = \neg \neg \neg \neg \neg A$
\item $A \wedgec B = \neg \neg ((\neg \neg A) \wedge (\neg \neg B))$
\item $A \veec B = \neg \neg ((\neg \neg A) \vee (\neg \neg B))$
\item $A \Rightarrowc B = \neg \neg ((\neg \neg A) \Rightarrow (\neg \neg B))$
\item $\fac x~A = \neg \neg (\fa x~(\neg \neg A))$
\item $\exc x~A = \neg \neg (\ex x~(\neg \neg A))$
\end{itemize}
\end{definition}
Notice that the propositions 
$\top \Leftrightarrow \topc$, 
$\bot \Leftrightarrow \botc$, and
$\neg A \Leftrightarrow \negc A$
where $A \Leftrightarrow B$ is defined as 
$(A \Rightarrow B) \wedge (B \Rightarrow A)$, 
have proofs. Thus, the symbols $\topc$, $\botc$, and $\negc$ could be just 
defined as $\top$, $\bot$, and $\neg$.

With this definition, neither the double negations in front of atomic
propositions nor those at the top of the proposition are missing. The
price to pay is to have four negations instead of two in many places,
but this is not harmful.

Yet, there is still a problem with the translation of atomic
propositions: as with any definition based translation, the atomic
proposition $P$ alone is translated as $P$ and not as $\neg \neg P$.
Thus, the property that a sequent $\Gamma \vdash A$ has a classical
proof if and only if the sequent $\|\Gamma\| \vdash \|A\|$ has a
constructive one only holds when $A$ is not atomic. For instance, the
sequent $P \wedgec Q \vdash P$, that is $\neg \neg ((\neg \neg P)
\wedge (\neg \neg Q)) \vdash P$, does not have a constructive proof.

A solution to this problem is to decompose hypothetical provability into
absolute provability and entailment. For absolute provability, 
the property that a sequent $\vdash A$ has a classical proof if and only if 
the sequent $\vdash \|A\|$ has a constructive one holds for all propositions,
because atomic propositions have no proof.
Thus, the sequent
$H_1, ..., H_n \vdash A$ has a classical proof if and only if the
sequent $\vdash \|H_1\| \Rightarrow^c ...\Rightarrow^c \|H_n\|
\Rightarrow^c \|A\|$ has a constructive one. This leads to a system where 
we have only one notion of absolute provability, but two notions of entailment:
``$A$ has a proof from the hypothesis $H$'' can either be understood
as ``$H \Rightarrow A$ has a proof'' or ``$H \Rightarrowc A$ has a
proof''.

\begin{figure}
\noindent\framebox{\parbox{\textwidth
}{
{
$$\begin{array}{cc}
\multicolumn{2}{c}{\irule{}
      {A \vdash A}
      {\mbox{axiom}}}
\\

\irule{\Gamma, A, A \vdash \Delta}
      {\Gamma, A \vdash \Delta}
      {\mbox{contr-l}}
&
~~~~~~~~~~~~~~~~~~~~

\irule{\Gamma \vdash A, A, \Delta}
      {\Gamma \vdash A, \Delta}
      {\mbox{contr-r}}\\

\irule{\Gamma \vdash \Delta}
      {\Gamma, A \vdash \Delta}
      {\mbox{weak-l}}
&
~~~~~~~~~~~~~~~~~~~~
\irule{\Gamma \vdash\Delta}
      {\Gamma \vdash A,\Delta}
      {\mbox{weak-r}}\\

&
~~~~~~~~~~~~~~~~~~~~
\irule{}
      {\Gamma \vdash \top, \Delta}
      {\mbox{$\top$-r}}\\

\irule{}
      {\Gamma, \bot \vdash \Delta}
      {\mbox{$\bot$-l}}\\

\irule{\Gamma \vdash A, \Delta}
      {\Gamma, \neg A \vdash  \Delta}
      {\mbox{$\neg$-l}}
&
~~~~~~~~~~~~~~~~~~~~
\irule{\Gamma, A \vdash \Delta}
      {\Gamma \vdash \neg A, \Delta}
      {\mbox{$\neg$-r}}\\

\irule{\Gamma, A, B \vdash \Delta}
      {\Gamma, A \wedge B \vdash  \Delta}
      {\mbox{$\wedge$-l}}
&
~~~~~~~~~~~~~~~~~~~~
\irule{\Gamma \vdash A, \Delta\hspace{0.5cm}\Gamma \vdash B, \Delta}
      {\Gamma \vdash A \wedge B, \Delta}
      {\mbox{$\wedge$-r}}\\

\irule{\Gamma, A \vdash \Delta \hspace{0.5cm} \Gamma, B \vdash \Delta}
      {\Gamma, A \vee B \vdash  \Delta}
      {\mbox{$\vee$-l}}
&
~~~~~~~~~~~~~~~~~~~~
\irule{\Gamma \vdash A, \Delta}
      {\Gamma \vdash A \vee B, \Delta}
      {\mbox{$\vee$-r}}\\
&
~~~~~~~~~~~~~~~~~~~~
\irule{\Gamma \vdash B, \Delta}
      {\Gamma \vdash A \vee B, \Delta}
      {\mbox{$\vee$-r}}\\

\irule{\Gamma \vdash A, \Delta \hspace{0.5cm} \Gamma, B \vdash \Delta}
      {\Gamma, A \Rightarrow B \vdash  \Delta}
      {\mbox{$\Rightarrow$-l}}
&
~~~~~~~~~~~~~~~~~~~~
\irule{\Gamma, A \vdash B, \Delta}
      {\Gamma \vdash  A \Rightarrow B, \Delta}
      {\mbox{$\Rightarrow$-r}}\\

\irule{\Gamma, (t/x)A \vdash \Delta}
      {\Gamma, \fa x~A \vdash \Delta}
      {\mbox{$\fa$-l}}
&
~~~~~~~~~~~~~~~~~~~~
\irule{\Gamma \vdash A, \Delta}
      {\Gamma \vdash \fa x~A, \Delta}
      {\mbox{$\fa$-r}}\\

\irule{\Gamma, A \vdash \Delta}
      {\Gamma, \ex x~A \vdash \Delta}
      {\mbox{$\ex$-l}}
&
~~~~~~~~~~~~~~~~~~~~
\irule{\Gamma \vdash (t/x)A, \Delta}
      {\Gamma \vdash \ex x~A, \Delta}
      {\mbox{$\ex$-r}}
\end{array}$$
}}}
\caption{Sequent calculus}
\label{classical}
\end{figure}

\begin{definition}[Classical and constructive provability]
Classical provability is defined by the cut free sequent calculus rules of
Figure \ref{classical}. 
We say that the proposition $A$ has a classical 
proof if the sequent $\vdash A$ does.

Constructive provability, our main 
notion of provability, is obtained by 
restricting to sequents with at most one conclusion. This requires a slight 
adaptation of the $\Rightarrow$-l rule
$$\irule{\Gamma \vdash A \hspace{0.5cm} \Gamma, B \vdash \Delta}
        {\Gamma, A \Rightarrow B \vdash  \Delta}
        {\mbox{$\Rightarrow$-l}}$$
We say that the proposition $A$ has a constructive 
proof if the sequent $\vdash A$ does.
\end{definition}

\begin{proposition}\label{direct}
If the proposition $\|A\|$ has a constructive proof,
then the proposition $A$ has a classical one.  
\end{proposition}

\noindent {\em Proof.} 
If the proposition 
$\|A\|$ has a constructive proof, then $\|A\|$
also has a classical proof. Hence, $\|A\|$ being classically
equivalent to $A$, $A$ also has a classical proof.

\smallskip

We now want to prove the converse: that if the proposition $A$ has a 
classical proof, then $\|A\|$ has a constructive proof.
To do so, we first introduce another translation where the top double
negation is removed, when there is one.
\begin{definition}[Light translation]~
\begin{itemize}
\item $|P| = P$,
\item $|\top| = \top$,
\item $|\bot| = \bot$,
\item $|\neg A| = \neg \neg \neg \|A\|$, 
\item $|A \wedge B| = (\neg \neg \|A\|) \wedge (\neg \neg \|B\|)$,
\item $|A \vee B| = (\neg \neg \|A\|) \vee (\neg \neg \|B\|)$,
\item $|A \Rightarrow B| = (\neg \neg \|A\|) \Rightarrow (\neg \neg \|B\|)$,
\item $|\fa x~A| = \fa x~(\neg \neg \|A\|)$,
\item $|\ex x~A| = \ex x~(\neg \neg \|A\|)$.
\end{itemize}
If $\Gamma = A_1, ..., A_n$ is a multiset of propositions, we write 
$|\Gamma|$ for the multiset $|A_1|, ..., |A_n|$ and 
$\neg |\Gamma|$ for the multiset $\neg |A_1|, ..., \neg |A_n|$. 
\end{definition}

\begin{proposition}\label{notnot}
If the proposition $A$ is atomic, then $\|A\| = |A|$, otherwise 
$\|A\| = \neg \neg |A|$. 
\end{proposition}

\noindent {\em Proof.}
By a case analysis on the form of the proposition $A$.

\begin{proposition}\label{oneandtheother}
If the sequent $\Gamma, |A| \vdash$ has a constructive proof, 
then so does the sequent $\Gamma, \|A\| \vdash$.
\end{proposition}

\noindent {\em Proof.} 
By Proposition \ref{notnot}, either $\|A\| = |A|$, or
$\|A\| = \neg \neg |A|$. In the first case the result is obvious, 
in the second, we build a proof of $\Gamma, \|A\| \vdash$ with 
a $\neg$-l rule, a a $\neg$-r rule, and the proof of $\Gamma, |A| \vdash$.

\begin{proposition}\label{key}
If the sequent $\Gamma \vdash \Delta$ has a classical
proof, then the sequent 
$|\Gamma|, \neg |\Delta| \vdash$
has a constructive one.
\end{proposition}

\noindent {\em Proof.}
By induction 
on the structure of the classical proof of the sequent
$\Gamma \vdash \Delta$. 
As all the cases are similar, we just give a few.

\begin{itemize}

\item If the last rule is the {\em axiom} rule, then $\Gamma =
\Gamma', A$ and $\Delta = \Delta', A$, and the sequent $|\Gamma'|,
|A|, \neg |A|, \neg |\Delta| \vdash$, that is $|\Gamma|, \neg
|\Delta| \vdash$, has a constructive proof.

\item If the last rule is the {\em $\Rightarrow$-l} rule, then
$\Gamma = \Gamma', A \Rightarrow B$ and by induction hypothesis,
the sequents $|\Gamma'|, \neg |A|, \neg |\Delta| \vdash$ and
$|\Gamma'|, |B|, \neg |\Delta| \vdash$ have constructive proofs,
thus the sequents $|\Gamma'|, \neg \|A\|, \neg |\Delta| \vdash$ and
$|\Gamma'|, \|B\|, \neg |\Delta| \vdash$ have constructive proofs,
thus, using Proposition \ref{oneandtheother}, 
the sequent $|\Gamma'|, \neg \neg \|A\|
\Rightarrow \neg \neg \|B\|, \neg |\Delta| \vdash$, 
that is
$|\Gamma|, \neg |\Delta| \vdash$, has a constructive proof.

\item If the last rule is the {\em $\Rightarrow$-r} rule, then 
$\Delta = \Delta',  A \Rightarrow B$ and by induction hypothesis, 
the sequent 
$|\Gamma|, |A|, \neg |B|, \neg |\Delta'| \vdash$
has a constructive proof, 
thus, using Proposition \ref{oneandtheother}, the sequent 
$|\Gamma|, \|A\|, \neg \|B\|, \neg |\Delta'| \vdash$
has a constructive proof, 
thus the sequent
$|\Gamma|, \neg (\neg \neg \|A\| \Rightarrow \neg \neg \|B\|), 
\neg |\Delta'| \vdash$,
that is  $|\Gamma|, \neg |\Delta| \vdash$, has a constructive proof.
\end{itemize}

\begin{proposition}\label{nonatom}
If the sequent $\Gamma \vdash A$ has a classical proof and $A$ is not 
an atomic proposition, then 
the sequent $\|\Gamma\| \vdash \|A\|$ has a constructive one.
\end{proposition}

\noindent {\em Proof.}
By Proposition \ref{key}, as the sequent $\Gamma \vdash A$ has a 
classical proof, the sequent $|\Gamma|, \neg |A| \vdash$ has a constructive 
one.
Thus, by Proposition \ref{oneandtheother}, the sequent $\|\Gamma\|, \neg |A| \vdash$ 
has a constructive proof, and the sequent 
$\|\Gamma\| \vdash \neg \neg |A|$ 
also.
By Proposition \ref{notnot}, as 
$A$ is not atomic, $\|A\| = \neg \neg |A|$. Thus, the sequent 
$\|\Gamma\| \vdash \|A\|$ has a constructive proof.

\begin{theorem}\label{main}
The proposition $A$ has a classical proof if and only if the proposition
$\|A\|$ has a constructive one.
\end{theorem}

\noindent {\em Proof.}
By Proposition \ref{direct}, if the proposition $\|A\|$ has a
constructive proof, then the proposition $A$ has a classical one.
Conversely, we prove that if the proposition $A$ has a classical
proof, then the proposition $\|A\|$ has a constructive one.  If $A$
is atomic, the proposition $A$ does not have a classical proof,
otherwise, by Proposition \ref{nonatom}, the proposition $\|A\|$ has
a constructive proof.

\begin{corollary}
The sequent $H_1, ..., H_n \vdash A$ has a classical proof 
if and only if the sequent $\vdash \|H_1\| \Rightarrowc ...
\|H_n\| \Rightarrowc \|A\|$ has a constructive one.
\end{corollary}

\noindent {\em Proof.}
The sequent $H_1, ..., H_n \vdash A$ has a classical proof
if and only if the sequent 
$\vdash H_1 \Rightarrow ...
H_n \Rightarrow A$ has one and, by Theorem \ref{main}, 
if and only if the sequent $\vdash \|H_1\| \Rightarrowc ...
\|H_n\| \Rightarrowc \|A\|$ has a constructive proof.

\smallskip

There is no equivalent of Theorem \ref{main} if we add double negations 
after the connectors only. For instance, the proposition
$P \vee \neg P$ has a classical proof, but the proposition
$\neg \neg P \vee \neg \neg \neg \neg \neg P$
has no constructive proof. O. Hermant \cite{Hermant} has proved that 
there is also no equivalent of Theorem \ref{main} if we add double 
negations 
before the connectors only. For instance, the proposition
$(\fa x~(P(x) \wedge Q)) \Rightarrow (\fa x~P(x))$ has a classical proof, 
but the 
proposition
$\neg \neg ((\neg \neg \fa x~\neg \neg (P(x) \wedge Q)) \Rightarrow 
(\neg \neg \fa x~P(x)))$ 
has no constructive proof.

Let $H1, ..., H_n$ be an axiomatization of mathematics with a finite
number of axioms, $H = H_1 \wedge ... \wedge H_n$ be their conjunction, 
and $A$ be a proposition. If the proposition
$H \Rightarrow A$ has a classical proof, then, by Theorem \ref{main}, 
the proposition $\|H\| \Rightarrowc \|A\|$ has a constructive one.  
Thus, in general, not only
the proposition $A$ must be formulated with classical connectives 
and quantifiers, but the axioms of the theory and the entailment relation 
also.

Using Proposition \ref{nonatom}, if $A$ is not an atomic proposition,
then the proposition $\|H\| \Rightarrow \|A\|$ has a constructive
proof.  In this case, the axioms of the theory must be formulated with
classical connectives and quantifiers, but the entailment relation
does not. 

In many cases, however, even the proposition $H \Rightarrow
\|A\|$ has a constructive proof.
For instance, consider the theory formed with the axiom $H$
``The union of two finite sets is finite'' 
$$\fa x \fa y~(F(x) \Rightarrow F(y) \Rightarrow F(x \cup y))$$
---or, as the cut rule is admissible in sequent calculus, any theory 
where this proposition has a proof---and let 
$A$ be the proposition 
``If the union of two sets is infinite then one of them is''
$$\fa a \fa b~((\neg F(a \cup b)) \Rightarrow (\neg F(a) \vee \neg F(b)))$$
which is, for instance,
at the heart of the proof of Bolzano-Weierstrass theorem, 
then the proposition $H \Rightarrow \|A\|$ has a constructive proof

$$\irule{
\irule{\irule{\irule{\irule{\irule{\irule{\irule{\irule{\irule{\irule{
\irule{\irule{\irule{\irule{}
                           {F(a), F(b) \vdash F(a)}
                           {\mbox{axiom}}
                     ~~~~~~~~~~~~~~~~~~~~~
                     \irule{\irule{}
                                  {F(a), F(b) \vdash F(b)}
                                  {\mbox{axiom}}
                            ~~~~~~~~~~
                            \irule{}
                                  {F(a \cup b), F(a), F(b) \vdash F(a \cup b)}
                                  {\mbox{axiom}}
                            }
                            {F(b) \Rightarrow F(a \cup b), F(a), F(b) \vdash F(a \cup b)}
                            {\mbox{$\Rightarrow$-l}}
                      }
                      {F(a) \Rightarrow F(b) \Rightarrow F(a \cup b), F(a), F(b) 
                       \vdash F(a \cup b)}
                      {\mbox{$\Rightarrow$-l}}
              }
              {H, F(a), F(b) \vdash F(a \cup b)}
              {\mbox{$\fa$-l, $\fa$-l}}
      }
      {H, \neg \neg (\negc F(a \cup b)),F(a), F(b) \vdash}
     {\mbox{$\neg$-l, $\neg$-r, $\neg$-l, $\neg$-r, $\neg$-l, $\neg$-r, $\neg$-l}}
                                                                      }
                                                                      {H, \neg \neg 
                                                                       (\negc F(a \cup
                                                                       b)), F(a) 
                                                                       \vdash \negc 
                                                                       F(b)}
                                                                      {\mbox{$\neg$-r, 
                                                                       $\neg$-l, 
                                                                       $\neg$-r,  
                                                                       $\neg$-l, 
                                                                       $\neg$-r}}
                                                               }
                                                               {H, \neg \neg (\negc 
                                                                F(a \cup b)), F(a) 
                                                                \vdash \negc F(a) 
                                                                \veec \negc F(b)}
                                                               {\mbox{$\neg$-l, 
                                                                $\neg$-r, $\vee$-r, 
                                                                $\neg$-r, $\neg$-l}}
                                                        }
                                                        {H, \neg \neg (\negc F(a 
                                                         \cup b)), \neg (\negc F(a) 
                                                         \veec \negc F(b)), F(a) 
                                                         \vdash}
                                                        {\mbox{$\neg$-l}}
                                                 }
                                                 {H, \neg \neg (\negc F(a \cup b)),
                                                  \neg (\negc F(a) \veec \negc F(b)) 
                                                  \vdash \negc F(a)}
                                                  {\mbox{$\neg$-r, $\neg$-l, $\neg$-r, 
                                                  $\neg$-l, $\neg$-r}}
                                          }
                                          {H, \neg \neg (\negc F(a \cup b)),
                                           \neg (\negc F(a) \veec \negc F(b)) \vdash
                                           \negc F(a) \veec \negc F(b)}
                                          {\mbox{$\neg$-r, $\neg$-l, $\vee$-r, 
                                           $\neg$-r, $\neg$-l}}
                                   }
                                   {H, \neg \neg (\negc F(a \cup b)),
                                    \neg (\negc F(a) \veec \negc F(b)), 
                                    \neg (\negc F(a) \veec \negc F(b)) \vdash}
                                  {\mbox{$\neg$-l}}
                            }
                            {H, \neg \neg (\negc F(a \cup b)),
                             \neg (\negc F(a) \veec \negc F(b)) \vdash}
                           {\mbox{contr-l}}
                     }
                     {H, \neg \neg (\negc F(a \cup b))
                          \vdash \neg \neg (\negc F(a) \veec \negc F(b))}
                     {\mbox{$\neg$-r}}
              }
              {H \vdash (\negc F(a \cup b)) \Rightarrowc (\negc F(a) \veec \negc F(b))}
              {\mbox{$\neg$-r, $\neg$-l, $\Rightarrow$-r}} 
        }
       {H \vdash \fac a \fac b~((\negc F(a \cup b)) \Rightarrowc (\negc F(a) 
        \veec \negc F(b)))}
       {\mbox{($\neg$-r, $\neg$-l, $\fa$-r, $\neg$-r, $\neg$-l)$^2$}}}
{\vdash H \Rightarrow \fac a \fac b~((\negc F(a \cup b)) \Rightarrowc (\negc F(a) 
        \veec \negc F(b)))}
{\mbox{$\Rightarrow$-r}}$$
In this case, even the proposition
$$H \Rightarrow \fa a \fa b~((\neg F(a \cup b)) \Rightarrow (\neg F(a) \veec \neg F(b)))$$
where the only classical connective is the disjunction, 
has a constructive proof.

Which mathematical results have a classical formulation that can be
proved from the axioms of constructive set theory or constructive type
theory and which require a classical formulation of these axioms 
and a classical notion of entailment remains to be investigated.

\section*{Acknowledgements}

The author wishes to thank Olivier Hermant and Sara Negri for many
helpful comments on a previous version of this paper.

\end{document}